\documentclass{article}
\usepackage{spconf,amsfonts,amsmath,amsthm,amssymb,graphicx}
\usepackage{adjustbox,lipsum,bm,soul}
\usepackage{cite}
\usepackage{color}
\usepackage{algorithm,algorithmic}
\DeclareMathOperator*{\argmin}{arg\,min}

\newcommand{\Complex}{{\mathbb{C}}}
\newcommand{\cs}{_{\text{\sf cs}}}
\newcommand{\admm}{_{\text{\sf ADMM}}}
\renewcommand{\vec}[1]{\ensuremath{\boldsymbol{#1}}}
\newcommand{\hvec}[1]{\ensuremath{\boldsymbol{\Hat{#1}}}}
\renewcommand{\eqref}[1]{Eq.~\ref{eq:#1}}
\newcommand{\algref}[1]{Algorithm~\ref{alg:#1}}

\DeclareMathOperator{\prox}{prox}
\newcommand{\defn}{\triangleq}

\newcommand{\figref}[1]{Fig.~\ref{fig:#1}}
\newcommand{\tabref}[1]{Table~\ref{tab:#1}}

\title{Free-breathing Cardiovascular MRI Using a Plug-and-Play Method with Learned Denoiser}
\name{Sizhuo Liu, Edward Reehorst, Philip Schniter, and Rizwan Ahmad\thanks{Corresponding author: Rizwan Ahmad (ahmad.46@osu.edu). This work was funded by NIH R01HL135489.}}
\address{The Ohio State University}
 \renewcommand{\vec}[1]{\ensuremath{\boldsymbol{#1}}}
\usepackage{color}

\begin{document}
 \abovedisplayskip=3pt 
 \belowdisplayskip=3pt 
 \abovedisplayshortskip=3pt 
 \belowdisplayshortskip=3pt 
 \arraycolsep=3pt

\maketitle
\begin{abstract}
Cardiac magnetic resonance imaging (CMR) is a noninvasive imaging modality that provides a comprehensive evaluation of the cardiovascular system. The clinical utility of CMR is hampered by long acquisition times, however. In this work, we propose and validate a plug-and-play (PnP) method for CMR reconstruction from undersampled multi-coil data. To fully exploit the rich image structure inherent in CMR, we pair the PnP framework with a deep learning (DL)-based denoiser that is trained using spatiotemporal patches from high-quality, breath-held cardiac cine images. The resulting ``PnP-DL'' method iterates over data consistency and denoising subroutines. We compare the reconstruction performance of PnP-DL to that of compressed sensing (CS) using eight breath-held and ten real-time (RT) free-breathing cardiac cine datasets. We find that, for breath-held datasets, PnP-DL offers more than one dB advantage over commonly used CS methods.  For RT free-breathing datasets, where ground truth is not available, PnP-DL receives higher scores in qualitative evaluation. The results highlight the potential of PnP-DL to accelerate RT CMR.

\end{abstract}
\section{Introduction}
\label{sec:intro}
Magnetic resonance imaging (MRI) is a well-established imaging modality that offers high soft-tissue contrast without the use of ionizing radiation. Cardiovascular magnetic resonance imaging (CMR) extends the application of MRI to yield static or dynamic images of the cardiovascular system. Cardiac cine, i.e., creating a movie of the beating heart, is one of the most common applications of CMR. Cardiac cine is typically performed under breath-holding conditions and requires regular heart rhythm. Under these conditions, the measured k-space data can be combined from several heartbeats, leading to a well-posed inverse problem. This approach, however, is not feasible for subjects who cannot hold their breath or are arrhythmic. For such patients, cardiac cine is performed in real-time (RT) and under free-breathing conditions. RT cine does not combine data across heartbeats, leading to high undersampling rates. 

To facilitate RT cine, several methods that combine multi-coil MRI and compressed sensing (CS) have been proposed. Selecting a sparsifying transform in CS, however, is non-trivial. Also, most commonly used transforms do not fully capture the rich structure of CMR images. More recently, deep learning (DL) methods have been shown to outperform CS methods. Some DL methods pose reconstruction as a ``de-aliasing'' problem, where the coil-combined aliased image is de-aliased using a convolutional neural network (CNN)~\cite{hyun2018deep}. These methods are computationally fast after the training phase, but ignore the multi-coil structure of the MRI data. 
In other methods, such as AUTOMAP~\cite{zhu2018image}, the training is used to learn the entire reconstruction process, from undersampled k-space to the final image, without any guidance from the forward model. The resulting CNN has fully connected layers, making it computationally prohibitive in practice. Also, such methods require extensive training to handle variations in the forward model. More recently, methods that explicitly use knowledge of the forward model~\cite{hammernik2018learning} have gained significant attention. In these methods, the training is guided by explicit inclusion of prototypical forward models.  For these methods, however, significant deviations between the forward models used for training versus testing can be problematic.

In this work, we recover CMR images from undersampled, multi-coil k-space data by iterating between data-fidelity enforcement and image denoising, i.e., via plug-and-play (PnP) recovery \cite{venkatakrishnan2013plug}. To fully exploit the rich spatiotemporal structure of CMR images, we use a DL-based denoiser trained specifically on CMR images. In the resulting ``PnP-DL'' method, training images are used to learn the denoiser, and these images are invariant to the forward model. Thus, the learned PnP-DL method is not biased by assumptions about the forward model, which can change significantly from training to testing. This unique feature of PnP-DL can lead to improved generalizability. We apply PnP-DL to reconstruct cardiac cine images and show that it outperforms traditional CS methods. More importantly, we demonstrate that PnP-DL can recover RT free-breathing cine images by using a denoiser trained on breath-held cine images. This capability is important because it is difficult, if not impossible, to obtain training data for RT free-breathing cine.

\section{Methods}
\subsection{PnP for Image Reconstruction}

We consider the problem of recovering an image $\vec{x}\in\Complex^N$ from noisy k-space measurements. Most CS methods recover $\vec{x}$ by solving an optimization problem of the form
\begin{align}
\hvec{x}\cs = \argmin_{\vec{x}} \big\{ f(\vec{x}) + \phi(\vec{x}) \big\}
\label{eq:CS},
\end{align}
where $f(\vec{x})$ is a (usually quadratic) data-fidelity term and $\phi(\vec{x})$ a convex, sparsity-promoting regularizer. A common algorithm to solve \eqref{CS} is the ``alternating directions method of multipliers'' (ADMM), which reformulates \eqref{CS} as
\begin{align}
\hvec{x}\admm = \argmin_{\vec{x}} \min_{\vec{v}}\big\{ f(\vec{x}) + \phi(\vec{v}) \big\}\quad \text{s.t.} \quad \vec{x}=\vec{v}
\label{eq:CS_ADMM},
\end{align}
and solves it using \algref{admm},
with $g(\vec{z})=\prox_{\nu\phi}(\vec{z})$ and
\begin{align}
\prox_{\nu\phi}(\vec{z}) 
&\defn \argmin_{\vec{x}} \left\{\phi(\vec{x}) + \tfrac{1}{2\nu}\|\vec{x}-\vec{z}\|_2^2\right\} 
\label{eq:prox},
\end{align}
which is known as the proximal operator.

\begin{algorithm}[h]
  \caption{The ADMM Algorithm}
  \label{alg:admm}
  \begin{algorithmic}[1]
    \REQUIRE $\nu>0, \vec{x}_0\in\Complex^N, \vec{u}_0\in\Complex^N$
	\FOR{$t=1,2,3,\dots$}
	\STATE{$\vec{v}_t = \argmin_{\vec{x}}\big\{ f(\vec{x}) + \frac{1}{2\nu}\|\vec{x}-(\vec{x}_{t-1}-\vec{u}_{t-1})\|_2^2 \big\}$}
	\label{line:admm_ls}
	\STATE{$\vec{x}_t = g(\vec{v}_t - \vec{u}_{t-1})$}
	\label{line:admm_denoise}
	\STATE{$\vec{u}_t = \vec{u}_{t-1} + (\vec{v}_{t}-\vec{x}_{t})$}
	\ENDFOR
	\RETURN{$\hvec{x}\admm \gets \vec{x}_t$}
  \end{algorithmic}
\end{algorithm}

From the Bayesian perspective, the 
proximal operator in \eqref{prox}
can be interpreted as the maximum a posteriori estimator of $\vec{x}$ from the noisy measurement $\vec{z} = \vec{x}+\vec{w}$, where $\vec{w}$ is $\nu$-variance additive white Gaussian noise and $\vec{x}$ has the probabilistic prior $p(\vec{x})\propto \exp(-\phi(\vec{x}))$.
Given this denoising interpretation, Bouman et al.~\cite{venkatakrishnan2013plug} replaced the proximal update in ADMM with a call to a sophisticated image denoising subroutine like BM3D and observed a dramatic improvement in the quality of image recovery. The approach was called ``plug-and-play'' (PnP) because essentially any image denoiser $g(\cdot)$ can be ``plugged into'' ADMM, including those based on DL.

\subsection{Learned Denoiser}
Compared to PnP methods that use generic denoisers (e.g., BM3D or BM4D), we propose to use a DL denoiser specifically trained for cardiac cine. First, it has been shown that application-specific DL denoisers can outperform generic denoisers~\cite{zhang2017beyond}. Second, CNN-based denoisers can be efficiently implemented on a GPU, unlike generic denoisers (e.g., BM4D), which can be prohibitively slow for the image sizes encountered in MRI. Third, most generic denoisers are designed only for real-valued 2D images, leaving limited options for complex-valued or higher-dimensional images. In contrast, DL denoisers can be trained and implemented for images in any domain or dimension. 

\begin{figure}
    \centering
    \includegraphics[width=8.6cm]{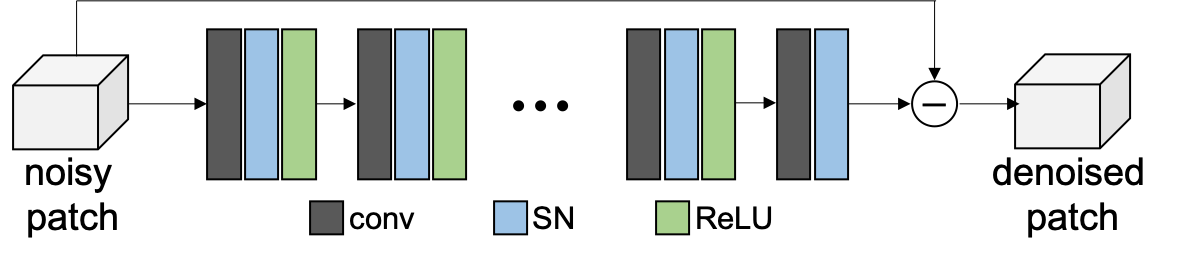}
    \caption{The spatiotemporal DL denoiser used in this work. Here, conv, SN, and ReLU stand for 3D convolution, spectral normalization, and rectified linear unit, respectively.}
    \label{fig:cnn}
\end{figure}

The DL-based denoiser used in this work has the architecture shown in \figref{cnn}. The denoiser is constructed with five 3D-convolution layers, each with 64 $3\times3\times3$ kernels.  Spectral normalization (SN) \cite{miyato2018spectral} was used to control the Lipschitz constant of each convolution layer and to provide some control of the overall Lipschitz constant of the network. We observed that the use of SN was critical for the stability of PnP~\cite{ryu2019plug}. The denoiser was trained using spatiotemporal patches extracted from high-quality, complex-valued cine images.

\section{Experiment and results}
\subsection{Experimental Setup}
For training, we acquired 50 fully sampled, breath-held cine datasets from eight healthy volunteers. To promote diversity in the training data, 28 short-axis, seven two-chamber, seven three-chamber, and eight four-chamber views were acquired. The reference images were reconstructed by taking the inverse FFT along phase and frequency encoding directions and combining the resulting coil images using the method \cite{walsh2000adaptive} by Walsh et al. Then, complex-valued i.i.d. Gaussian noise was added to these ``noise-free'' reference images to simulate noisy images with an SNR of 26 dB. For training, we cropped each image into patches of size $55\times55\times15$, with the last dimension representing time. The resulting noisy and noise-free patches were assigned as input and target output when training the DL denosier. We fed the real and imaginary parts into separate channels, set the minibatch size to four, and used the ADAM optimizer with a learning rate of $10^{-4}$ over 500 epochs on an NVIDIA GPU (GeForce RTX 2080 Ti).

\begin{figure}[t]
    \includegraphics[width = 8.6cm,trim=0 10 0 10,clip]{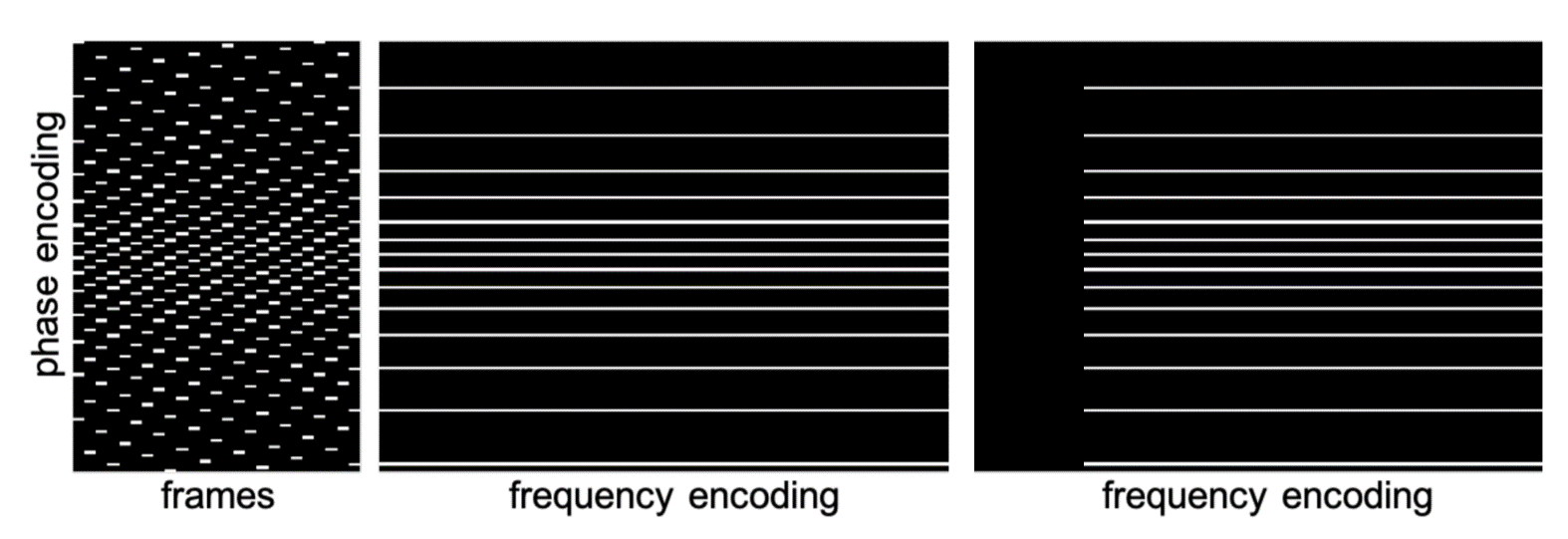}
    \centering
    \caption{A representative sampling pattern. Left: phase encoding and frame (time) dimensions shown for one frequency encoding location. Middle: phase encoding and frequency encoding dimensions shown for the last frame. Right: same as the middle column but with asymmetric echo, which was only used in prospective undersampling.}
    \label{fig:samp}
\end{figure} 

For testing, nine breath-held datasets were collected from four different healthy volunteers, with four in the short-axis view, two in the two-chamber view, two in the four-chamber view, and one in the axial view. Note, the axial view was not included in the training datasets. All datasets were retrospectively undersampled at three different acceleration rates: $R$ = 6, 8, and 10. We also collected ten RT free-breathing datasets with prospective undersampling at $R$ = 9. Pseudo-random Cartesian masks were used to perform undersampling; an example is shown in \figref{samp}. Before reconstruction, all datasets were compressed to 12 virtual coils for faster processing. For PnP-DL, the reconstruction followed \algref{admm}, with $f(\vec{x})$ representing the SENSE-based forward model. The coil sensitivity maps were estimated using the method \cite{walsh2000adaptive} by Walsh et al. For the recruitment and consent of human subjects used in this study, the ethical approval was given by an Internal Review Board (2005H0124) at The Ohio State University.

\begin{figure}[t]
    \includegraphics[width=8.2cm]{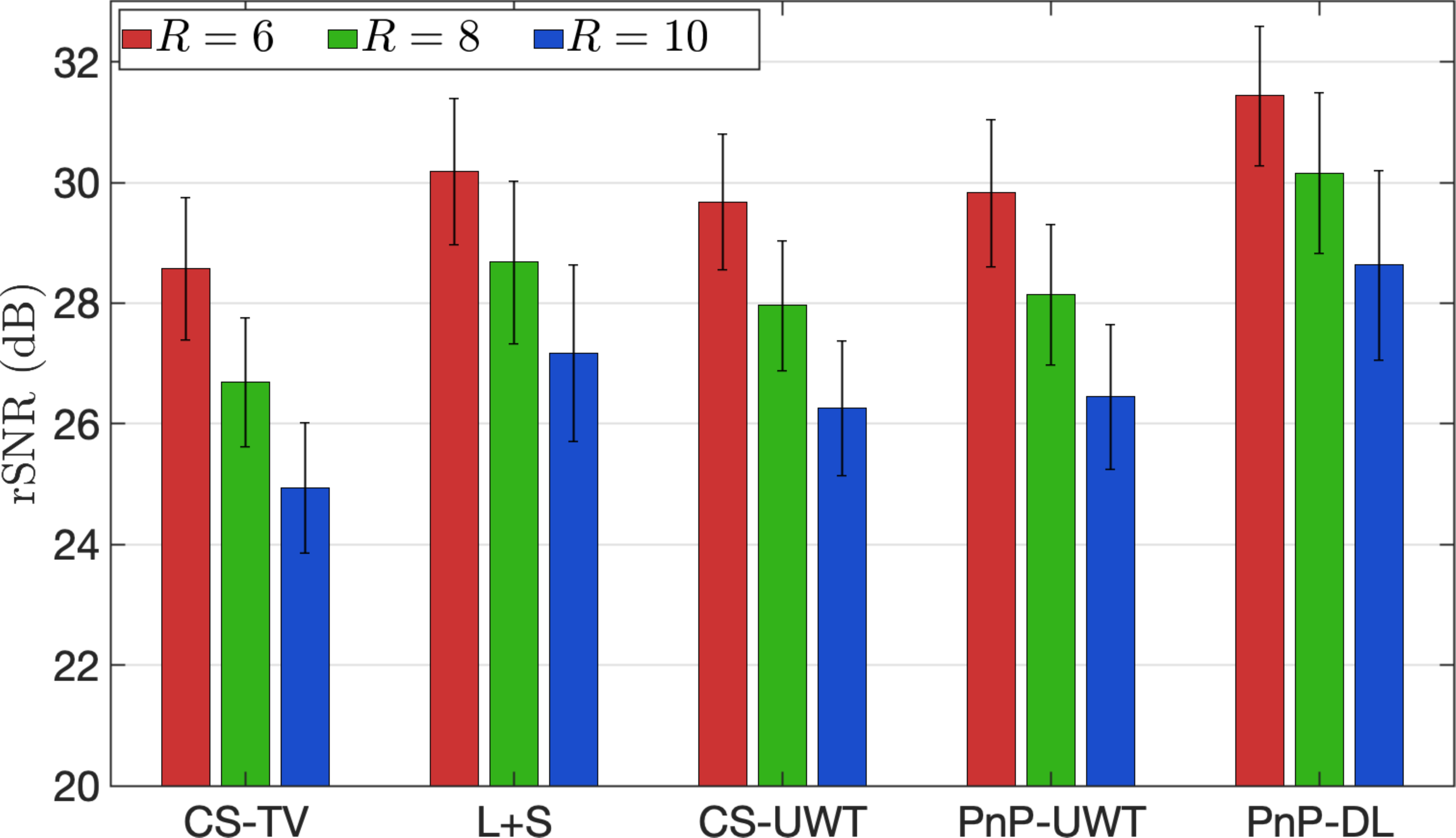}
    \centering
    \caption{Reconstruction SNR (rSNR) for breath-held datasets retrospectively undersampled at $R$ = 6, 8, and 10.}
    \label{fig:retro_rnsr}
\end{figure}

\begin{table}[t]
\smallskip
\centering
\begin{adjustbox}{width=0.45\textwidth}
\begin{tabular}{|l|l|l|l|l|l|}
\hline
 & \textbf{CS-TV} & \textbf{L+S} & \textbf{CS-UWT} & \textbf{PnP-UWT} & \textbf{PnP-DL}\\ \hline \hline
$R=6$  & 30.6    & 32.1 & 31.6 & 31.9 & \bf{33.5} \\ \hline
$R=8$  & 29.0    & 31.1 & 30.2 & 30.5 & \bf{32.5} \\ \hline
$R=10$ & 27.1    & 29.4 & 28.5 & 28.8 & \bf{31.3} \\ \hline
\end{tabular}
\end{adjustbox}
\caption{Reconstruction SNR (rSNR) for the retrospectively undersampled dataset collected in the axial view.}
\label{tab:rsnr}
\end{table}

\subsection{Results from Breath-held Data}
In the study with retrospective undersampling, we evaluated the reconstructed images using reconstruction SNR (rSNR), defined as $20\log_{10}(\|\vec{x}\|_2/\|\vec{x}-\hvec{x}\|_2)$, where $\hvec{x}$ and $\vec{x}$ represent reconstructed and reference cine images, respectively. The results from PnP-DL and PnP-UWT were compared to those from CS-TV~\cite{lustig2007sparse}, CS-UWT~\cite{ting2017fast}, and L+S~\cite{otazo2015lps}, which represent CS with spatiotemporal finite difference as the sparsifying transform, CS with undecimated wavelet transform (UWT) as the sparsifying transform, and low-rank plus sparse reconstruction, respectively. For PnP-UWT, denoising was accomplished by soft-thresholding the UWT coefficients. Therefore, CS-UWT and PnP-UWT solve the same optimization problem, but CS-UWT used bFISTA~\cite{ting2017fast} while PnP-UWT used ADMM from \algref{admm}. PnP-DL also used ADMM, but with the trained DL-based denoiser shown in~\figref{cnn}. 

\figref{retro_rnsr} shows rSNR values aggregated over eight datasets and for three acceleration rates. The dataset collected in the the axial view was not included. 
As expected, the performances of PnP-UWT and CS-UWT were similar. The average advantage of PnP-DL over L+S (second best) was 1.3~dB, 1.5~dB, and 1.5~dB for $R$ = 6, 8, and 10, respectively. 

\begin{figure*}[t]
    \includegraphics[width = 17.4cm]{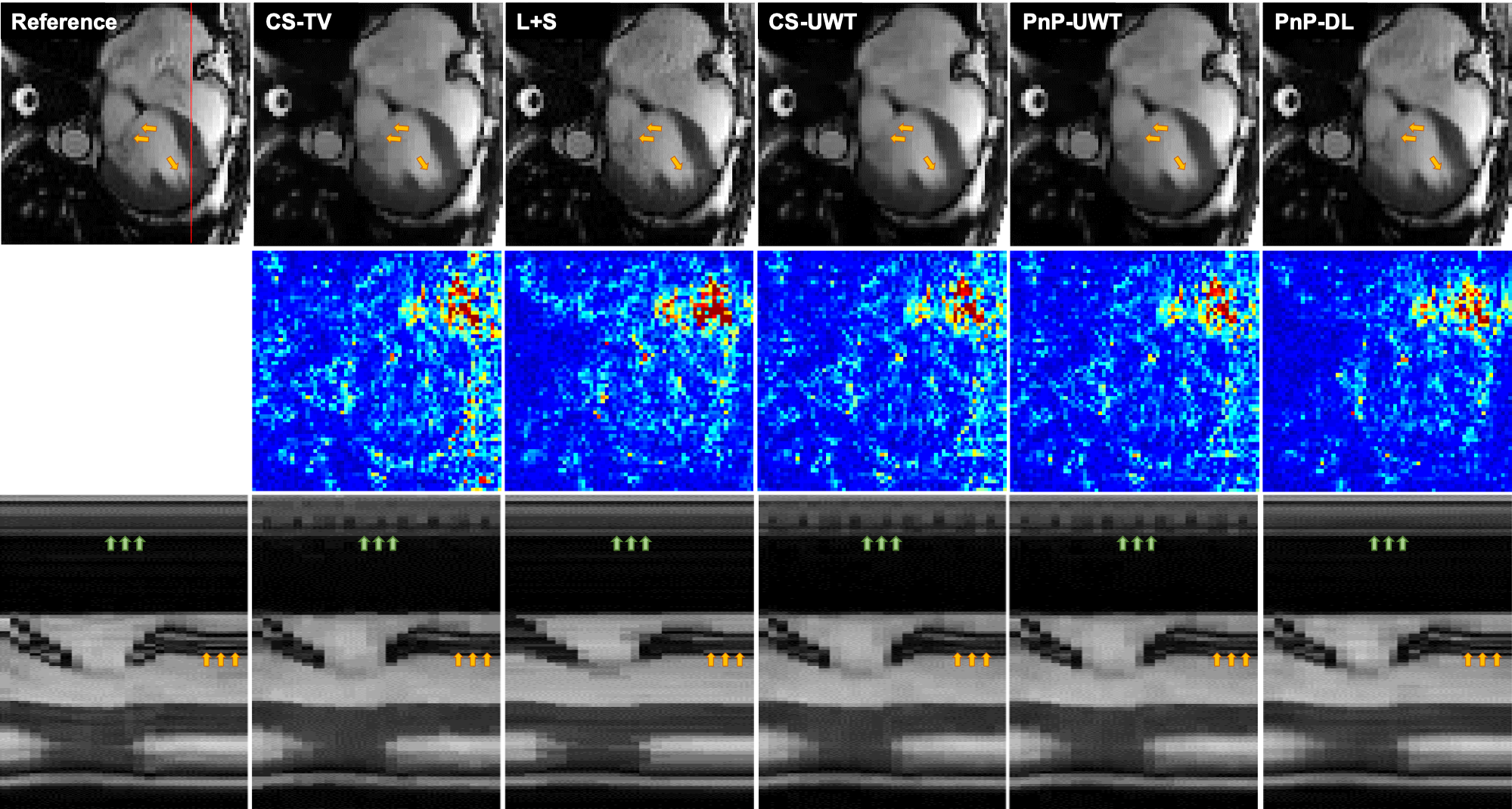}
    \centering
    \caption{Top row: A representative frame from the retrospectively undersampled dataset collected in the axial view. The arrows highlight the details that are better preserved in PnP-DL. Middle row: Error map with five-fold amplification. Bottom row: Temporal profile along the red line shown on the Reference image in the top row.}
    \label{fig:axial}
\end{figure*}

\tabref{rsnr} compares rSNR of the ninth dataset, which was collected in the axial view. Note, the axial view was not included when training the denoiser. Even for this ``unseen'' view, PnP-DL maintains its advantage over the CS methods. The superiority of PnP-DL is also evident from the example recoveries in \figref{axial}, which show that PnP-DL did better in preserving the mitral valve and papillary muscles (yellow arrows), as well as in suppressing artifacts (green arrows).

\subsection{Results from Free-breathing Data}
\begin{figure}[h!]
    \includegraphics[width = 8.2cm]{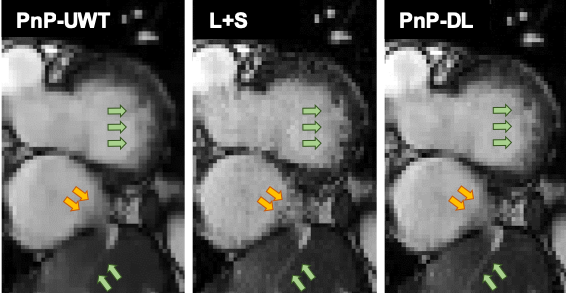}
    \centering
    \caption{A representative frame from an RT dataset.}
    \label{fig:pros}
\end{figure}

In the study with prospective undersampling, we reconstructed ten RT cine datasets collected under free-breathing conditions. Each dataset was reconstructed using PnP-DL, PnP-UWT, and L+S. To evaluate the reconstructed images, each cine series was blindly reviewed by an expert, with seven years of experience in CMR, who assigned a qualitative score on a five-point Likert scale (1: non-diagnostic, 2: poor, 3: adequate, 4: good, 5: excellent). L+S, PnP-UWT, and PnP-DL received average scores of 3.2, 3.5, and 3.9, respectively. Also, in nine out of ten cases, the PnP-DL recoveries were deemed best among the three reconstruction methods. As seen in \figref{pros}, which shows a representative frame from one of the RT free-breathing datasets, PnP-DL was able to preserve details lost with PnP-UWT (green arrows) and was more effective in removing the noise that is evident in the L+S recoveries (yellow arrows).

\section{Conclusions}
In this work, we proposed and validated a reconstruction method for real-time cardiac MRI. The method utilizes an application-specific deep-learning denoiser within the plug-and-play framework. The learning process, which entails training a denoiser on spatiotemporal image patches, is totally decoupled from the forward model. Our preliminary results suggest that PnP-DL outperforms CS in both quantitative and qualitative assessments. More importantly, PnP-DL was also effective in reconstructing real-time, free-breathing cine images even when the denoiser was trained using only breath-held cine images.



\bibliographystyle{IEEEbib}
\bibliography{root.bib}

\end{document}